\documentclass[final,5p,times]{elsarticle}

\usepackage{graphicx}
\usepackage{amssymb}
\usepackage{amsmath}
\usepackage{bm}
\usepackage{lineno}
\usepackage{caption}
\usepackage{subcaption}
\usepackage[mathscr]{euscript}

\usepackage{amsthm}
\usepackage{color}
\usepackage{natbib}
\usepackage{titlesec}
\usepackage{xcolor}
\usepackage{float}
\usepackage{hyperref}
\usepackage{hypernat}
\usepackage{ulem}
\usepackage{tabularx}

\begin{document}

\begin{frontmatter}

\title{neuSIM4: A comprehensive GEANT4 based neutron simulation code}
\author[a,b]{J. Park\corref{cor}} \ead{jhpark95@korea.ac.kr}
\author[c,d]{F. C. E. Teh}
\author[c,d]{M. B. Tsang\corref{cor}} \ead{tsang@frib.msu.edu}
\author[c,e]{K. W. Brown}
\author[f]{Z. Chajecki}
\author[a,b]{B. Hong}
\author[c]{T. Lokotko}
\author[c,d]{W. G. Lynch}
\author[c,d]{J. Wieske}
\author[c,d]{K. Zhu}

\cortext[cor]{Corresponding author}

\address[a]{Center for Extreme Nuclear Matters (CENuM), Korea University, Seoul 02841, Republic of Korea}
\address[b]{Department of Physics, Korea University, Seoul 02841, Republic of Korea}
\address[c]{Facility for Rare Isotope Beams (FRIB), Michigan State University, East Lansing, MI 48824 USA}
\address[d]{Department of Physics and Astronomy, Michigan State University, East Lansing, MI 48824 USA}
\address[e]{Department of Chemistry, Michigan State University, East Lansing, MI 48824 USA}
\address[f]{Department of Physics, Western Michigan University, Kalamazoo, MI 49008 USA}

\begin{abstract}
A new \textbf{neu}tron \textbf{SIM}ulation program based on the versatile GEANT\textbf{4} toolkit, \textbf{neuSIM4}, 
has been developed to describe interactions of neutrons in the NE213 liquid scintillator from 0.1 to 3000 MeV. \textbf{neuSIM4} is designed to accommodate complicated modern detector geometry setups with multiple scintillator detectors, each of which can be outfitted with more than one photo-multiplier. 
To address a broad spectrum of neutron energies, two new neutron interaction physics models, KSCIN and NxQMD, have been implemented in GEANT4. For neutrons with energy below 110 MeV, we incorporate a total of eleven neutron induced reaction channels on hydrogen and carbon nuclei, including nine carbon inelastic reaction channels, into KSCIN. Beyond 110 MeV, we implement a neutron induced reaction model, NxQMD, in GEANT4. We use its results as reference to evaluate other neutron-interaction physics models in GEANT4. We find that results from an existing cascade physics model (INCL) in GEANT4 agree very well with the results from NxQMD, and results from both codes agree with new and existing light response data. To connect KSCIN to NxQMD or INCL, we introduce a transition region where the contribution of neuSIM4 linearly decreases with corresponding increased contributions from NxQMD or INCL. To demonstrate the application of the new code, we simulate the light response and performance of a 2 $\times$ 2 m$^{2}$ neutron detector wall array consisting of 25 2m-long scintillation bars. We are able to compare the predicted light response functions to the shape of the experimental response functions and calculate the efficiency of the neutron detector array for neutron energies up to 200 MeV. These simulation results will be pivotal for understanding the performance of modern neutron arrays with intricate geometries, especially in the measurements of neutron energy spectra in heavy-ion reactions. 

\end{abstract}


\end{frontmatter}

\section{Introduction} \label{sec:intro}

Neutrons are integral components of all nuclei except protons. As a result, emission of neutrons are ubiquitous in all nucleus-nucleus collisions. Detecting neutrons is essential for gaining insights into the structure of neutron-rich nuclei~\cite{erler-1}, understanding the nature of asymmetric nuclear matter~\cite{lynch-1, russotto-1}, and unraveling the astrophysical processes that power nucleosynthesis~\cite{schatz-1}. To understand the nature of asymmetric nuclear matter, it is important to determine the density dependence of the symmetry energy. The symmetry energy is a term embedded in the nuclear equation of state. As it increases with the neutron-proton asymmetry of the system, it becomes very important in understanding the properties of neutron stars. A systematic comparison of proton and neutron spectra from nuclear matter with differing neutron-proton compositions is one of the experimental observables that can be used to constrain the symmetry energy. A major challenge in this effort lies in the determination of the neutron detection efficiency which is often less than 10\% with significant uncertainty, in contrast to the nearly 100\% intrinsic charged particle detection efficiency achievable with silicon and scintillator detectors, e.g. the High Resolution Array (HiRA) telescope~\cite{wallace-1} and MUST2~\cite{pollacco-1}.
Consequently, the detection and analysis of neutron data are more complicated and difficult to understand than that of the charged particles and therefore demands comprehensive understanding of the detector performance using simulations~\cite{shim-1}. Presently, there is no comprehensive neutron interaction simulation code designed to cover a wide range of energy, especially above 20 MeV, while offering enough flexibility to incorporate complex detector geometries in setups. 

Various neutron simulation codes such as the GEANT3~\cite{brun-1}, FLUKA~\cite{fluka-1}, NRESP~\cite{dietze-1}, SCINFUL~\cite{dickens-1}, and GEANT4~\cite{agostinelli-1} toolkit based MENATE$\_$R~\cite{kohley-1} have been used to calculate light response functions and detection efficiencies for neutron detectors. However, the valid energy ranges of NRESP are limited to 20 MeV, and the MENATE$\_$R code does not include sufficient neutron induced reaction channels~\cite{coupland-1} to accurately describe neutron interactions. On the other hand, SCINFUL gives reliable results of response functions and detection efficiencies for neutron detectors with neutron energies up to 110 MeV. To extend the energy beyond that, a quantum molecular dynamic transport model (QMD) has been used to describe neutrons emitted at high energy. It is important to note that QMD is a generic name for quantum molecular dynamic models developed to simulate heavy ion collisions. Over the years, many QMD codes which employ different approximations and techniques to simulate the nucleus-nucleus collisions have been developed. The calculated results of QMD codes may vary from each other~\cite{hermann-1}. In this context, we label the QMD code used in SCINFUL-QMD as NxQMD as explained in later section. In this paper, we use the more accurate name SCINFUL-NxQMD.

The SCINFUL-NxQMD~\cite{satoh-1, satoh-com} code was developed by the Japan Atomic Energy Agency (JAEA) to simulate the response of NE213 scintillation neutron detectors across a broad spectrum of neutron energies up to 3 GeV. However, it is limited to model only cylindrical scintillators, each of which has one photomultiplier (PMT) affixed at the back. While small cylindrical neutron detectors were prevalent in experiments conducted before the 1990s, this fixed geometry severely restricts its application on simulating the performance of contemporary neutron arrays consisting of many detectors, often arranged in sophisticated and diverse geometries. 

In this paper, we describe our work to develop a neutron simulation code, neuSIM4, that can provide reliable response functions of a neutron detection array with full flexibility to accommodate different detector geometries over a wide range of neutron energies. We use the framework of GEANT4 which is a flexible simulation toolkit that allows to easily model configuration of complicated detector geometries. Furthermore, new physics models can be implemented easily. To ensure accuracy of the neutron simulations over a wide range of energies, the interface of neuSIM4 limits users to incorporate only well-tested physics models described in this work. 

\begin{figure}[!t]
\begin{center}
\includegraphics[width=0.90\linewidth]{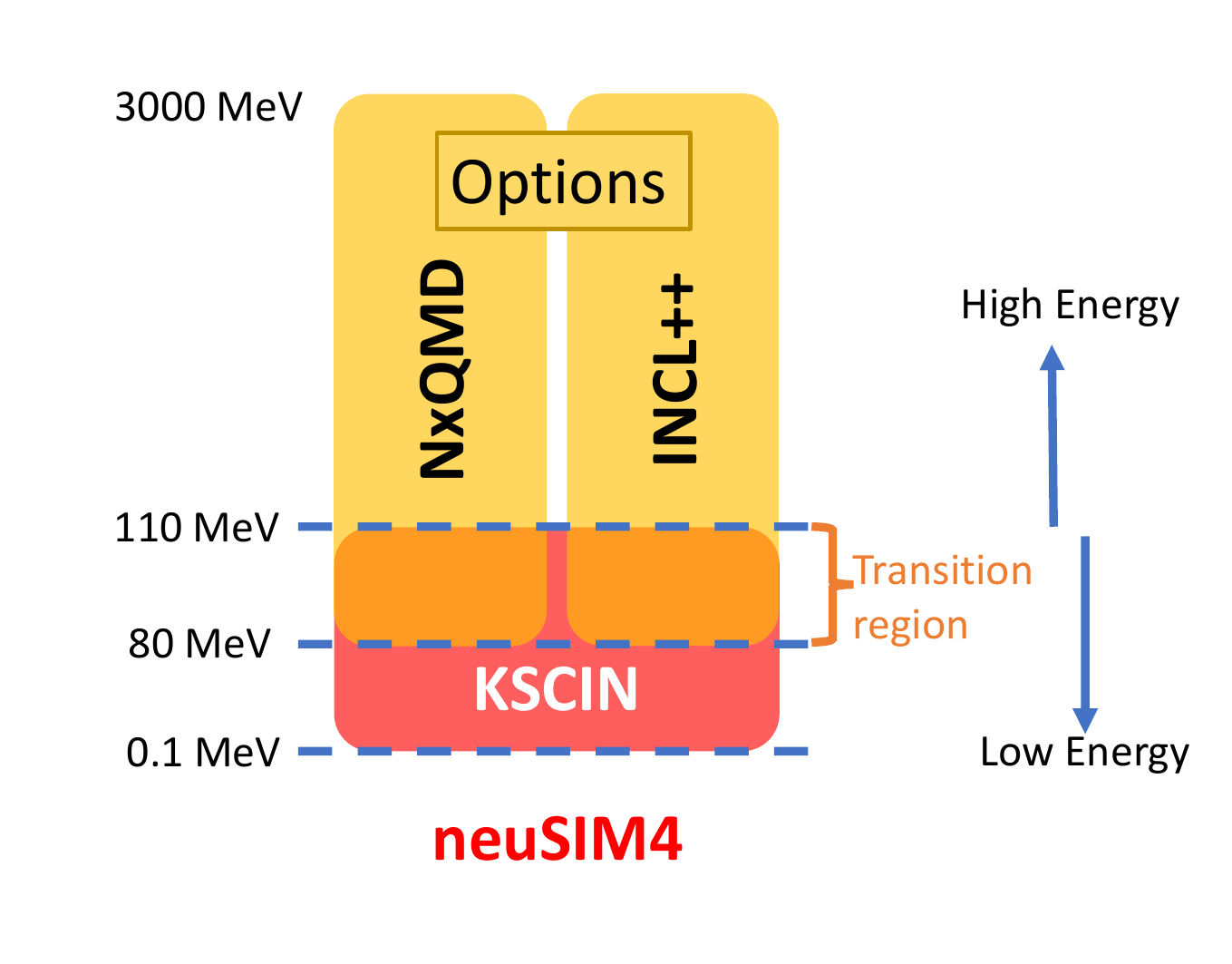}
\end{center}
\caption{Physics models used in neuSIM4. KSCIN is used for the low energy of neutrons. Both NxQMD and INCL++ model can be incorporated to simulate the interactions of high energy of neutron as user option. The transition region between 80 to 100 MeV is also schematically illustrated. The energy labels are not to scale. They are used mainly for illustration purpose.}
\label{fig:flow_chart}
\end{figure}

The components of neuSIM4 are illustrated schematically in the flow chart shown in Fig.~\ref{fig:flow_chart}. To simplify discussions, we designate low energy as below 110 MeV and high energy as above 110 MeV unless otherwise noted. A major part of our work focuses on the implementation of two new physics models, KSCIN (SCINFUL developed in Korea) for low energy neutrons and NxQMD (a quantum molecular dynamic transport model that describes nucleon induced collisions on heavy nuclei developed by the JAEA ~\cite{niita-1}) for high energy neutrons. As explained in later section, the C++ version of INCL~\cite{mancusi-1} which is an existing physics model in GEANT4 performs similarly as NxQMD and is much less CPU intensive. It is implemented as an option in neuSIM4. We use the conventions of neuSIM4(NxQMD) and neuSIM4(INCL) to distinguish the two options. To minimize the effects of the discontinuity in the calculated quantity at the energy when one switches from KSCIN to NxQMD or INCL, we linearly decrease contributions from KSCIN and simultaneously increase the contributions from high energy physics models as described in Ref.~\cite{satoh-4} in the region of 80 MeV to 110 MeV. Different transition regions have been explored and the uncertainties of such a procedure are small. Nonetheless, an option is provided for users to change the transition region.

This article is organized as follows. After this brief introduction, Sec.~\ref{sec:scinful-qmd} describes the implementation of KSCIN and NxQMD into GEANT4. Sec.~\ref{sec:light-output-in-G4} details techniques used in determination of the light output response function. In Sec.~\ref{sec:scinful-geant4} we validate our codes by comparing our results with SCINFUL-NxQMD calculations for the exact cylindrical detector geometry as published in~\cite{satoh-1, satoh-com}. In the same section, we also compare results from the new code to results from different neutron physics models found in GEANT4. To demonstrate the flexibility of neuSIM4, it is used to simulate the neutron response and efficiency of a large area neutron wall array (LANA) located at the Facility for Rare Isotope Beams (FRIB). The configuration of LANA is described in Sec.~\ref{sec:lana}. Comparison of response functions to new data obtained from the collisions of $^{48}$Ca + $^{64}$Ni at 140 MeV/u~\cite{fanurs-thesis}, and determination of neutron detection efficiencies for LANA, are given in Sec.~\ref{sec:lana-sim} and Sec.~\ref{sec:lana-sim-eff}. Finally, a summary is given in Sec.~\ref{sec:sum} that includes some future perspectives.

\section{New Physics Models} \label{sec:scinful-qmd}

Simulations of neutron scattering in detector materials depend on the energy of the neutrons. At low energy, the simulations depend on the number of included reaction channels and accuracy of database libraries in the models~\cite{antolkovic-1, subramanian-1, mcmillan-1, kellogg-1}. For high energy neutrons, only total scattering cross-section data for n+H and n+C reactions are available. Reaction models are needed to predict the emission of neutrons, mostly from sequential decays of excited fragments created during n+$^{12}$C-nuclear collisions. It therefore becomes necessary to develop a combination of low energy and high energy models to simulate the performance of neutrons over a wide range of neutron energies. Our code is modeled after the successful neutron simulation code, SCINFUL-NxQMD, which is a marriage of two models: SCINFUL for low energy and NxQMD, for high energy neutrons~\cite{satoh-com}. While the original codes were written in the FORTRAN language~\cite{dickens-1, satoh-com}, KSCIN and NxQMD are written in C++. 

\begin{table}
    \centering
    \begin{tabular} {ccc}
    \hline
     Reaction    & Q-value (MeV) & Threshold (MeV)\\
     \hline
    $^{1}$H(n,n)$^{1}$H              & 0 & 0\\
    $^{12}$C(n,n)$^{12}$C            & 0 & 0\\
    $^{12}$C(n,n$'$ )$^{12}$C$^{*}$  & -4.433  & 4.812\\
    $^{12}$C(n,$\alpha$)$^{9}$Be     & -5.71   & 6.186\\
    $^{12}$C(n,n$'$)3$\alpha$        & -7.656  & 8.4\\
    $^{12}$C(n,p)$^{12}$B            & -12.613 & 13.665\\
    $^{12}$C(n,d)$^{11}$B            & -13.732 & 15.25\\
    $^{12}$C(n,pn$'$)$^{11}$B        & -15.957 & 17.35\\
    $^{12}$C(n,2n)$^{11}$C           & -18.72  & 20.3\\
    $^{12}$C(n,t)$^{10}$B            & -18.93  & 21.5\\
    $^{12}$C(n,$^{3}$He)$^{10}$Be    & -19.47  & 22.0\\
    \hline
    \end{tabular}
    \caption{Neutron induced reaction channels compiled in~\cite{satoh-1} and included in KSCIN.}
    \label{tab:reaction_list}
\end{table}

\subsection{KSCIN} \label{sec:scinful-in-G4}
The original version of the low energy code, SCINFUL, was developed at Oak Ridge National Laboratory (ORNL) in 1988 for fast neutrons with energies from 0.1 MeV up to 80 MeV~\cite{dickens-1}. Organic scintillators primarily consist of hydrogen and carbon atoms. SCINFUL distinguishes itself from other neutron simulation codes through the care of the SCINFUL developers in evaluating their database for many neutron induced carbon reactions. KSCIN follows the same philosophy. The list of 11 neutron induced reactions in KSCIN is shown in the top panel of Fig.~\ref{fig:scinful_qmd}. More details about these reactions can be found in Table~\ref{tab:reaction_list}. The elastic scattering of neutrons on protons, $^{1}$H(n,n)$^{1}$H and carbon $^{12}$C(n,n)$^{12}$C, are represented by the black solid and black dotted lines in the bottom left panel of Fig.~\ref{fig:scinful_qmd}. These two channels have the largest cross-sections. However $^{1}$H(n,n)$^{1}$H decreases sharply with energy such that above 10 MeV, neutron induced reactions on $^{12}$C become the dominant reaction channel. 

For the inelastic scattering of $^{12}$C(n,n')$^{12}$C, the scattered $^{12}$C nucleus with excitation energy below 8 MeV deexcites mainly by emitting gamma rays. For fragments with excitation energy above the particle emission thresholds, light charged particles such as p, d, t, $^{3}$He and $^{4}$He can be emitted (see Table~\ref{tab:reaction_list}). The probabilities of each of the neutron induced reactions listed in Table~\ref{tab:reaction_list} are determined using the compiled cross-sections shown in Fig.~\ref{fig:scinful_qmd} and incorporated into KSCIN. The code then proceeds to calculate the excitation energy of the residual fragments. This is estimated by finding the energy difference between the initial and final states of the reaction as described in~\cite{dickens-1}. If the residual nucleus has sufficient excitation energy, it will either decay further into a daughter nucleus or decay to its ground state by emitting gamma rays. Detailed descriptions on how the deposited energy is converted to light are given in Sec.~\ref{sec:light-output-in-G4}.

\subsection{NxQMD} \label{sec:NxQMD-in-G4}

\begin{figure}[t!]
\begin{center}
\includegraphics[width=0.98\linewidth]{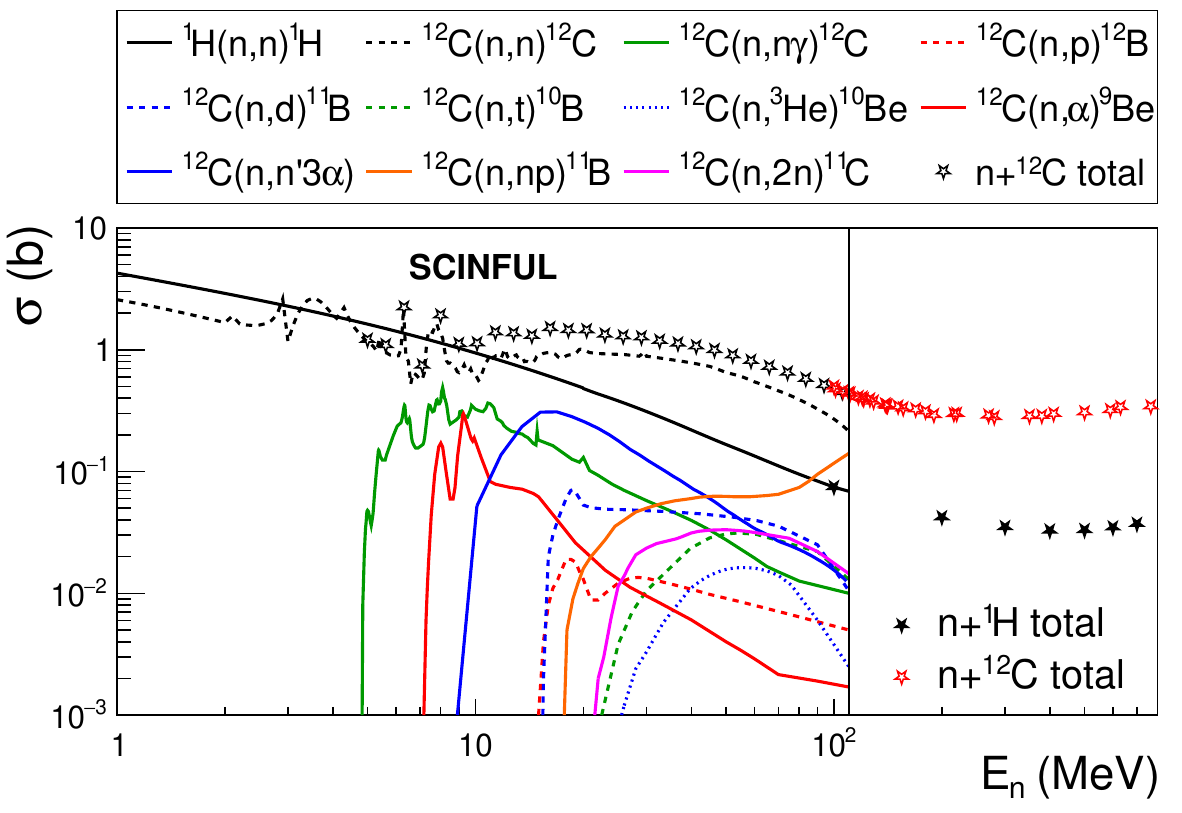}
\end{center}
\caption{Neutron induced reaction cross-sections as a function of neutron energy included in neuSIM4. The bottom panels show the cross-sections of various reaction channels from~\cite{satoh-1, satoh-com}. The evaluated total cross-sections for n+C and n+H from Ref.~\cite{chiba-1, monning-1} are plotted as red open stars and black solid stars respectively. The top panel contains the legend for different reactions which are listed in Table~\ref{tab:reaction_list} and plotted in the bottom left panel. The database for individual C induced reactions are compiled and evaluated by the JAEA with neutron energy up to 110 MeV. This figure is modified from Figs. 4 and 11 in~\cite{satoh-com}.}
\label{fig:scinful_qmd}
\end{figure}

For neutrons with energy above 110 MeV, data from individual neutron induced $^{12}$C reactions does not exist. Only the total cross-sections of n+H~\cite{chiba-1} and n+$^{12}$C~\cite{ monning-1} reactions are available. These are shown in the right side of Fig.~\ref{fig:scinful_qmd}, red stars for n+$^{12}$C and black stars for n+H. Note that the low energy and high energy data overlap at energy around 100 MeV. At high energy, the n+$^{12}$C cross-section is ten times higher than that for the n+H reaction. 

To calculate the n+$^{12}$C reaction channels, a reaction model is needed. We implement a quantum molecular model, NxQMD, as the high neutron energy physics model in GEANT4. NxQMD extends the validity of simulated neutron energy up to 3 GeV~\cite{satoh-1, satoh-com}. 

In QMD models, single nucleon particle states are represented by Gaussian wave functions, and a nucleus (many-body particle state) is represented by the product of these single particle wave functions for the nucleons within it. The QMD model numerically simulates the motions of these wave functions in nucleus-nucleus collisions. In NxQMD ~\cite{niita-1}, a freeze-out time is set to 100 fm/$c$, which is equivalent to 3.33 $\times$ 10$^{-22}$ seconds after the start of the collisions. At freeze-out, the calculation is stopped and the evolution of the positions and momenta of individual nucleons from the start of each simulated collision to its freeze-out time are then determined. Using a fragment recognition algorithm, which imposes a radius of 4 fm to chain any two nucleons to be part of a fragment as described in~\cite{niita-1}. At the end of the chain, the composition of the final fragments and their excitation energies are then determined. The decays of these excited primary fragments are calculated with a Statistical Decay Model (SDM). The final emitted particles from the SDM mainly consist of protons, neutrons, deuterons, tritons, and the helium isotopes $^{3}$He and $^{4}$He~\cite{niita-1}. The conversion of light charged particle energy into light output is described in the next section.

\section{Response function of the detector} \label{sec:light-output-in-G4}

To construct the response function of the detector, for every scattered incident neutron, the charged particles generated by the neutron induced reactions propagate in the scintillator material and keep depositing their energies in each step along the track. Then, the deposited energy is converted into light in units of MeV$_{\rm{ee}}$ using the empirical formulas developed in~\cite{satoh-1,satoh-com,satoh-3} for different particles. In the case that the light charged particle may have a large enough kinetic energy to escape the detector, the light nuclei (including protons) make less light than the fully stopped particles. To account for this effect, the light output $L_{n}$ was calculated for the $n$th step, using the following equation: 
\begin{equation}
L_{n}(E_{n}) = L(E_{n}) - L(E_{n} - E^{dep}_{n}),
\label{eq:light_output_step}
\end{equation}
where $L_{n}(E_{n})$ is the light output of the charged particle at the $n$th step, $L(E_{n})$ is the empirical light output formula as a function of kinetic energy of the charged particle introduced in~\cite{satoh-1,satoh-com,satoh-3}, $E_{n}$ is the kinetic energy of the charged particle in the beginning of the $n$th step and $E^{dep}_{n}$ is the energy deposited in the detector in the same step. 

Finally, the response function which depends on the neutron energy as well as the geometry of the detector, was constructed with the total light output $L_{tot}$, for which $L_{n}(E_{n})$ was summed up from $n$=1 to the index number corresponding to the moment either the particle is fully stopped in the detector or escapes the detector. In the last stage of the simulation, all the light output values produced by the neutron induced reactions are stored event by event and normalized by the incident number of neutrons. 
The corresponding neutron efficiency is determined by integrating the full response function from a given light output threshold to infinity (in practice up to 200 MeV$_{\rm{ee}}$). Accurate comparisons require that the light output thresholds must be set to be identical for both simulation and the data.

\begin{figure}[t!]
\begin{center}
\includegraphics[width=0.93\linewidth]{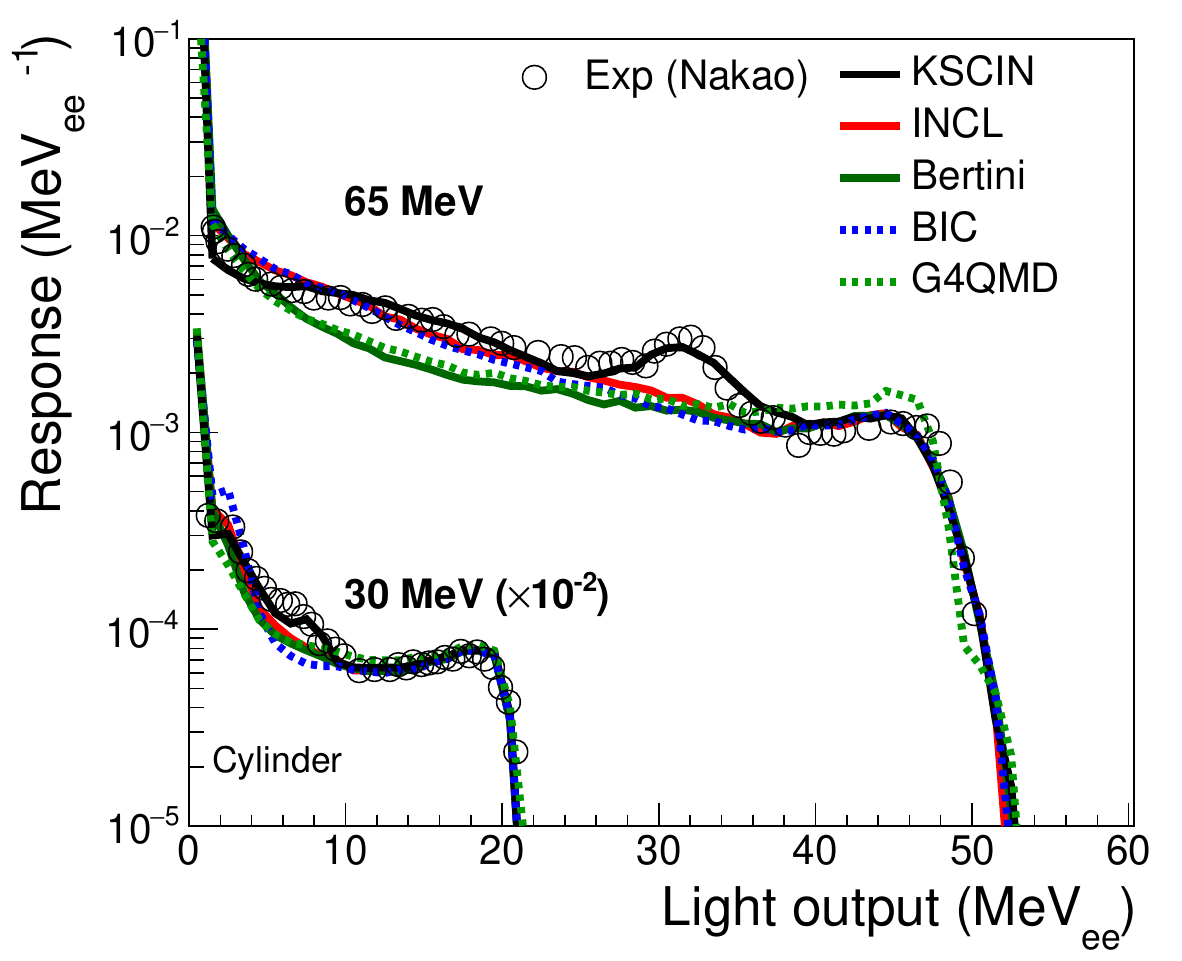}
\end{center}
\caption{Response functions for 30 and 65 MeV neutrons produced by different cascade, QMD and KSCIN physics models. Symbols are experimental data from Ref~\cite{nakao-1}. The $^{12}$C(n,d) peak can only be reproduced by KSCIN. The detector is modeled as a cylinder with a diameter of 12.7 cm and length of 12.7 cm.}
\label{fig:low_response}
\end{figure}

\begin{figure}[t!]
\begin{center}
\includegraphics[width=0.93\linewidth]{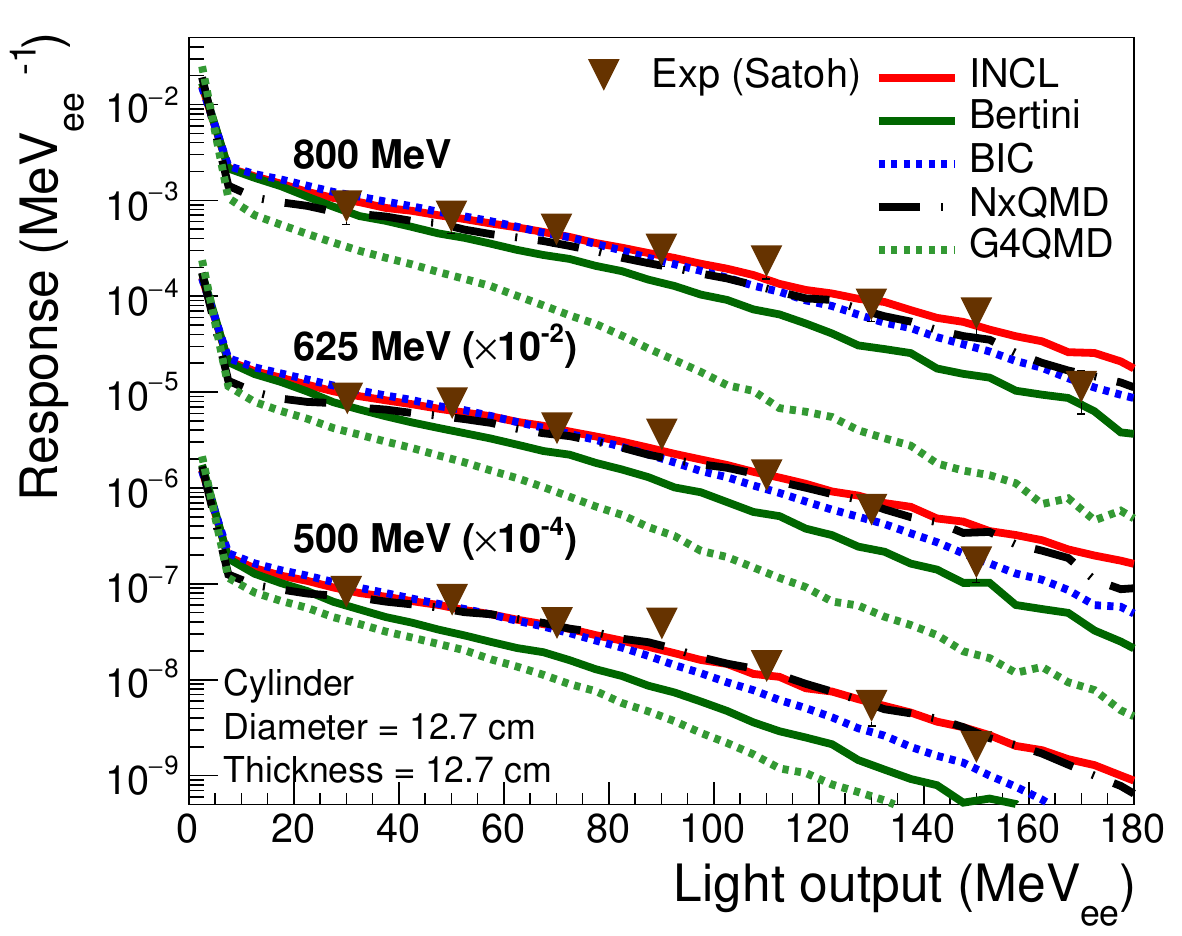}
\includegraphics[width=0.93\linewidth]{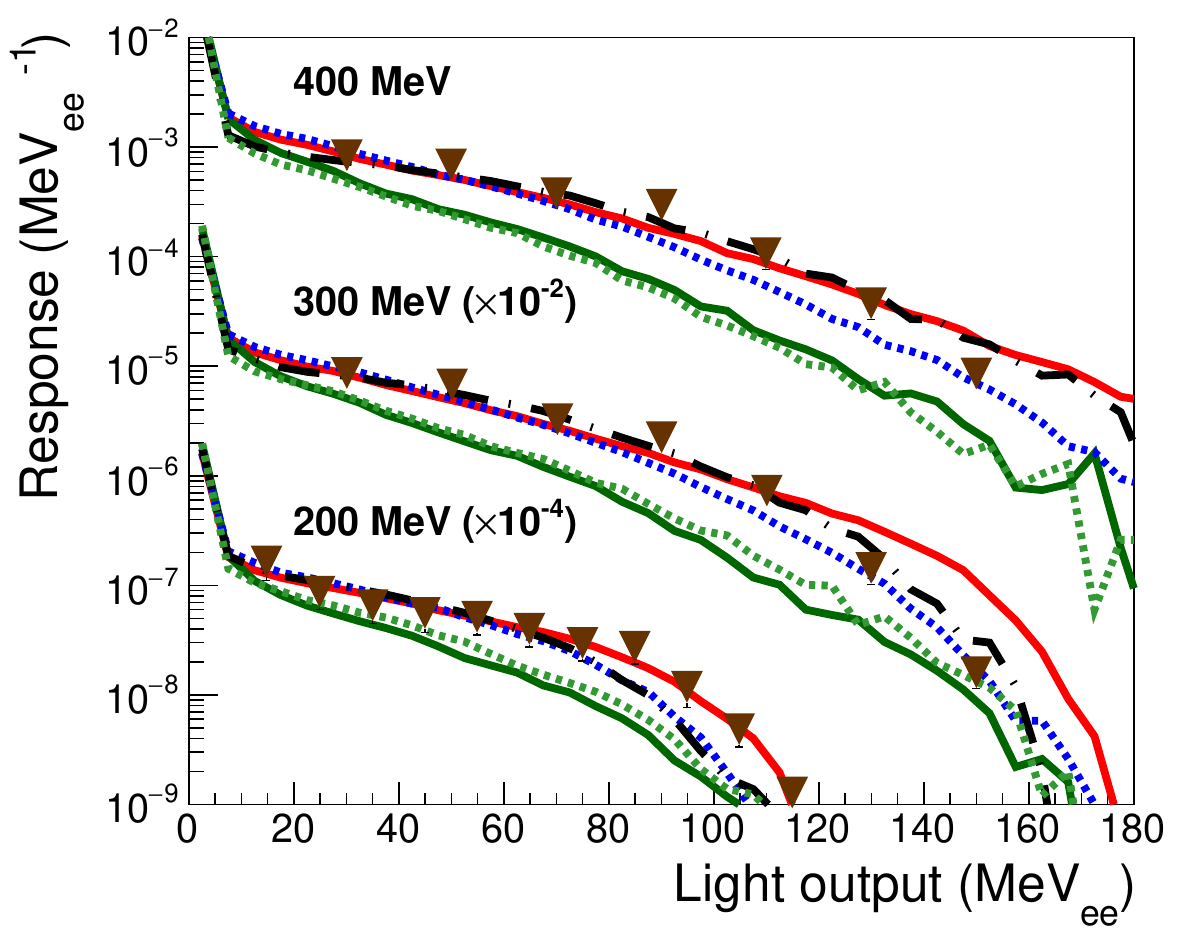}
\end{center}
\caption{Response functions for high-energy neutrons with different physics models. The detector is modeled as a cylinder with a diameter of 12.7 cm and length of 12.7 cm. Symbols are experimental data from Ref.~\cite{satoh-3}}
\label{fig:cylinder_resp}
\end{figure}

\section{Comparison of the performance of physics models on neutron scattering in GEANT4} \label{sec:scinful-geant4}

The recent update of the physics model G4ParticleHP in GEANT4, using the evaluated neutron data library G4NDL~\cite{mendoza-1, thulliez-1} coupled with NRESP mode~\cite{garcia-1}, improves greatly the accuracy of the simulations for neutron energy below 20 MeV. The introduction of the KSCIN physics model into GEANT4 represents a major advancement in neutron detector simulations that simulations of neutron interactions can now be performed accurately up to 110 MeV.  

In GEANT4, a range of cascade models are available, including Li\`ege Intranuclear Cascade (INCL)~\cite{boudard-1, boudard-2, mancusi-1}, Binary Cascade (BIC)~\cite{folger-1}, and Bertini intranuclear cascade (Bertini)~\cite{gurthrie-1}. In addition, another QMD model called G4QMD~\cite{koi-1} is also available as a physics model. Comparison of INCL to Bertini, BIC and G4QMD have been studied in~\cite{mancusi-1} to describe the fragment and light particle cross-sections as well as angular distributions from light charged particle induced reactions. The results from these models differ widely. In general INCL performs better than the other codes.

For verification of the new code, and comparison of the results from GEANT4 physics models to that of SCINFUL-NxQMD (which only supports cylindrical detector geometry), we adopt a cylindrical NE213 detector with both a diameter and length of 12.7 cm as modeled in~\cite{satoh-1, satoh-com}. A point source is located 4.5 m away from the detector endcap center, and the neutrons are uniformly sent to the detector in a cone. 

Since cascade and QMD models are constructed mainly to describe reactions at high energy, without consideration of nuclear structures, their deficiencies at low energy can be clearly demonstrated by comparing their results to data in Fig.~\ref{fig:low_response}. As they don't include the cross-sections of the different inelastic channels listed in Fig.~\ref{fig:scinful_qmd}, they cannot reproduce the peak from the $^{12}$C(n,d) reaction in the experimental data. In contrast, KSCIN reproduces the peak and the fall-off of the response function.

In Fig.~\ref{fig:cylinder_resp}, we compare the light response functions from different cascade and QMD codes with neutron energy from 200 to 800 MeV where experimental data are available~\cite{satoh-2}. As expected, the NxQMD results fit the data very well~\cite{satoh-1, satoh-com}. On the other hand, G4QMD, fits the data the worst. The cascade model INCL (red solid curves) also does a good job in describing the data, especially at 200 MeV neutron energy where the performance of INCL is better than NxQMD. However the performance of those two reverses at 300 MeV. Below 500 MeV (lower panel), results from Bertini (dark green solid lines) are similar to those from G4QMD (green dotted curves), and both of their results are lower than BIC, INCL and NxQMD. The comparison results are consistent with previous study~\cite{mancusi-1}. Compared to  QMD models which describe time evolution of many-body dynamics, cascade models mainly describe nucleon-nucleon collisions and neglect detailed many-body correlations. The simplified and approximate treatments undertaken by cascade codes result in computation speeds that are about 100 times faster than QMD codes. Based on our comparisons, INCL is as good as NxQMD. As described earlier, NxQMD is developed to specifically describe nucleon induced reactions on nucleus ~\cite{niita-1} while INCL is originally developed to describe nucleon induced reactions. During development, both codes were applied to describe similar proton induced reaction data~\cite{niita-1,boudard-1,boudard-2}. This might explain why the calculations of the light output from neutron induced reactions from both codes are similar.

\begin{figure}[t!]
\begin{center}
\includegraphics[width=0.80\linewidth]{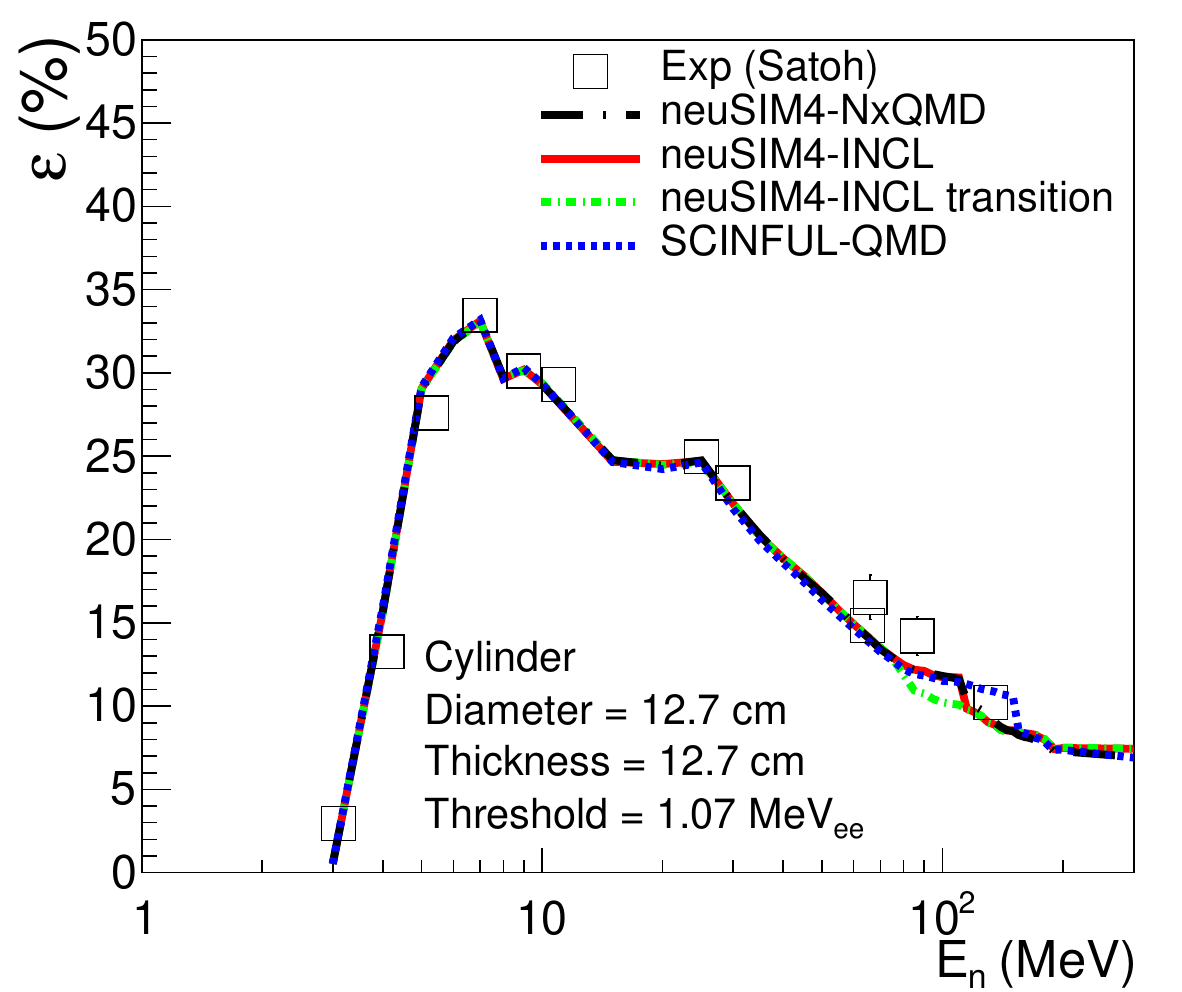}
\includegraphics[width=0.80\linewidth]{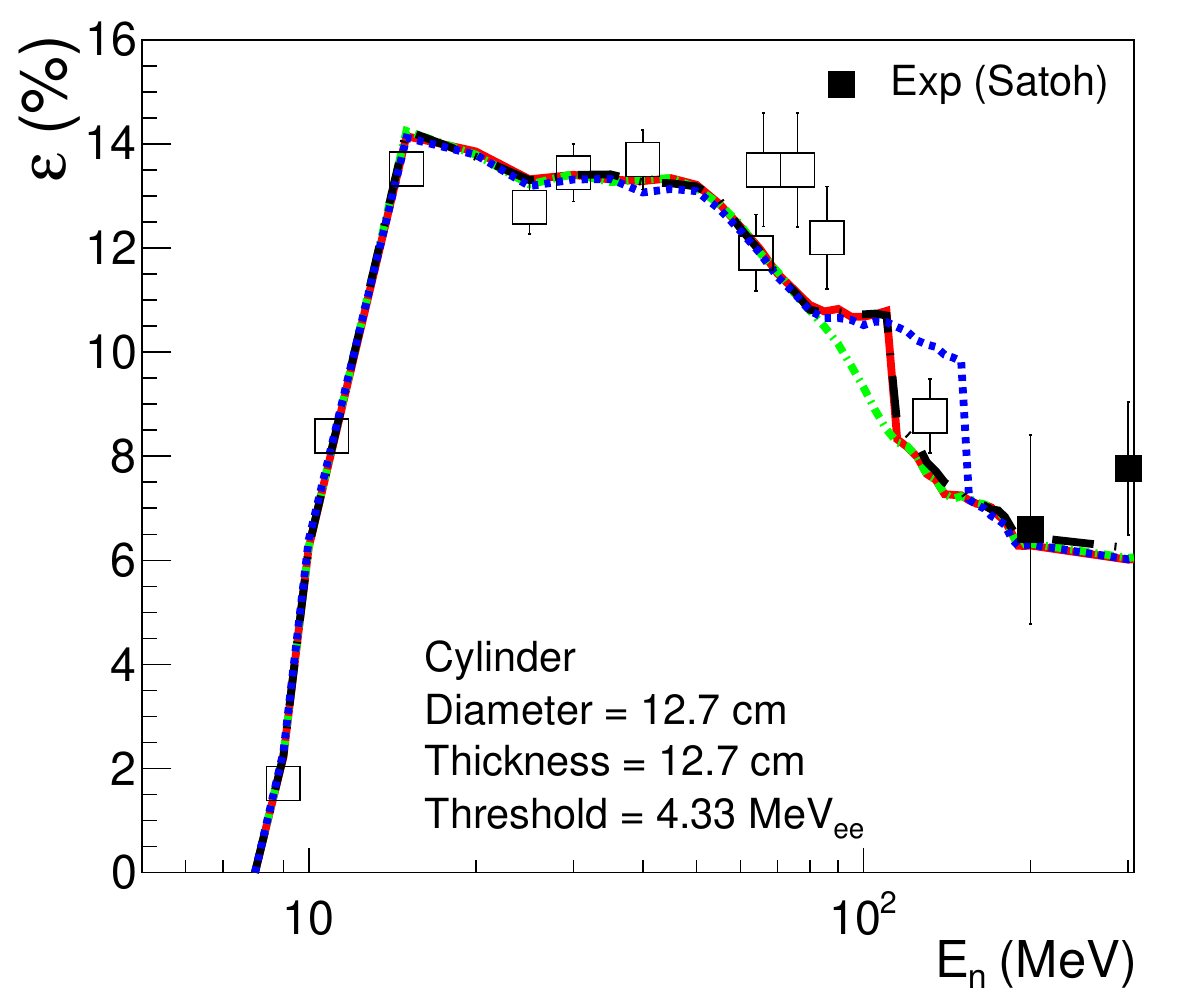}
\end{center}
\caption{Neutron detection efficiencies as a function of the neutron energy. The experimental data are taken from Ref.~\cite{satoh-4} and~\cite{satoh-2}. Light output threshold values of 1.07 and 4.33 MeVee have been applied in the top and bottom panels, respectively. The reference results from SCINFUL-NxQMD are represented by blue dotted lines. The discontinuity at 150 MeV neutron energy ($E_{n}$) is due to switching from SCINFUL to NxQMD at that energy. Similarly, we also note the discontinuities in neuSIM4(INCL) (red solid lines) or neuSIM4(NxQMD) (black dot-dashed line) at 110 MeV which is the limit of the reaction database shown in Fig.~\ref{fig:scinful_qmd}. The green dot-dashed line indicates the transition from 80 to 110 MeV for neuSIM4(INCL).}
\label{fig:cylinder_eff}
\end{figure}

By integrating the response function from light output threshold values of 1.07 and 4.33 MeV$_{\rm{ee}}$ to infinity, neutron detection efficiencies for the cylindrical detector from neuSIM4(NxQMD), neuSIM4(INCL), and the original SCINFUL-NxQMD are shown in Fig.~\ref{fig:cylinder_eff} as a function of neutron energy $E_{n}$. The experimental data taken from~\cite{satoh-4} are compared to simulation results. The discontinuity in the efficiency curve at 150 MeV from the original SCINFUL-NxQMD code (blue dotted lines) is due to abruptly switching from SCINFUL to ``NxQMD" at $E_{n}$=150 MeV~\cite{satoh-com}. By design, we switch from low energy code to high energy code at 110 MeV. Fig.~\ref{fig:cylinder_eff} shows the effect of the discontinuity at 110 MeV for neuSIM4(INCL) (red solid curve) or neuSIM4(NxQMD) (black dot-dashed curve). The green dot-dashed line shows the result when the transition occurs smoothly from 80 to 110 MeV as described earlier and in~\cite{satoh-4}. 
The uncertainties using different regions for transition are small especially when very low energy light output thresholds are applied as shown in the upper panel of Fig.~\ref{fig:cylinder_eff}. 

\section{Application of the new code to a neutron wall} %

To verify the simulation results of our codes in real experimental configuration, we used the data obtained from a large area neutron wall array, LANA, located at FRIB. LANA has been used to detect neutrons with energies up to 200 MeV from nuclear collisions~\cite{coupland-1, zecher-1, fanurs-thesis, zhu-1}. By measuring time of flight, it provides kinetic energy information for detected neutrons with high resolution.

\subsection{Configuration of LANA} \label{sec:lana}

LANA consists of two large neutron walls and one veto wall. Each neutron wall has cross-sectional dimensions of 2 $\times$ 2 m$^{2}$, and consists of 25 cuboids encapsulated by a Pyrex glass wall with thickness of 3.18 mm. Each 2-m long cuboid has cross-sectional dimensions of 7.62(height) $\times$ 6.35(depth) cm$^{2}$ filled with NE213 liquid scintillator. Two photo multipliers (PMT) are coupled to the two ends of each bar for light collection from the scintillator. NE213 is chosen for its good separation of neutrons from gammas using the pulse-shape discrimination (PSD) method~\cite{teh-1}. The upstream faces of the two neutron walls for LANA are located at 441.6 and 517.5 cm, respectively, from the nominal target position. The center of LANA is placed at a polar angle of 39.4$^{\circ}$ with respect to the beam direction to avoid direct exposure to the beam. 

Since neutron detectors cannot distinguish charged particles and neutrons, a charged particle veto wall was constructed and placed in front of the neutron walls to improve neutron detection by eliminating the abundant charged particle background. To completely cover LANA, the veto wall consists of 25 plastic scintillator bars each with area 250(height) $\times$ 9.4(width) cm$^{2}$ and thickness of 1 cm. Like the neutron wall bars, a PMT is coupled at each end of every veto wall bar to collect light. Neighboring bars overlap by 3 mm to eliminate any charged particles escaping through gaps. The veto wall modules are arranged vertically so that they form a 2D grid with the horizontal neutron wall bars. The hit position in an individual neutron/veto wall bar is determined by using the time difference between the light signals measured at each end.

\subsection{Simulated light output response for the Large Area Neutron Array, LANA} \label{sec:lana-sim}

\begin{figure}[t!]
\begin{center}
\includegraphics[width=0.85\linewidth]{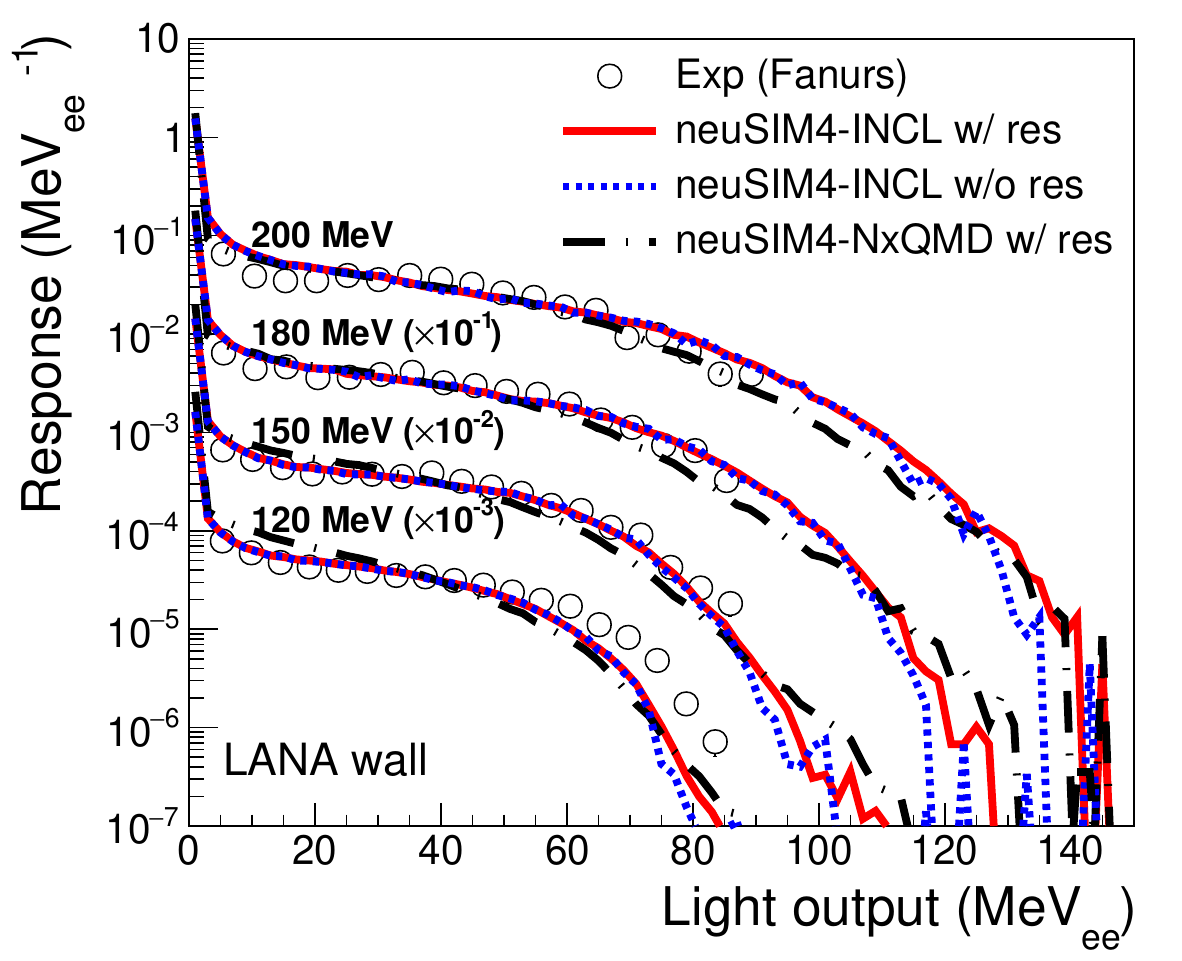}
\includegraphics[width=0.85\linewidth]{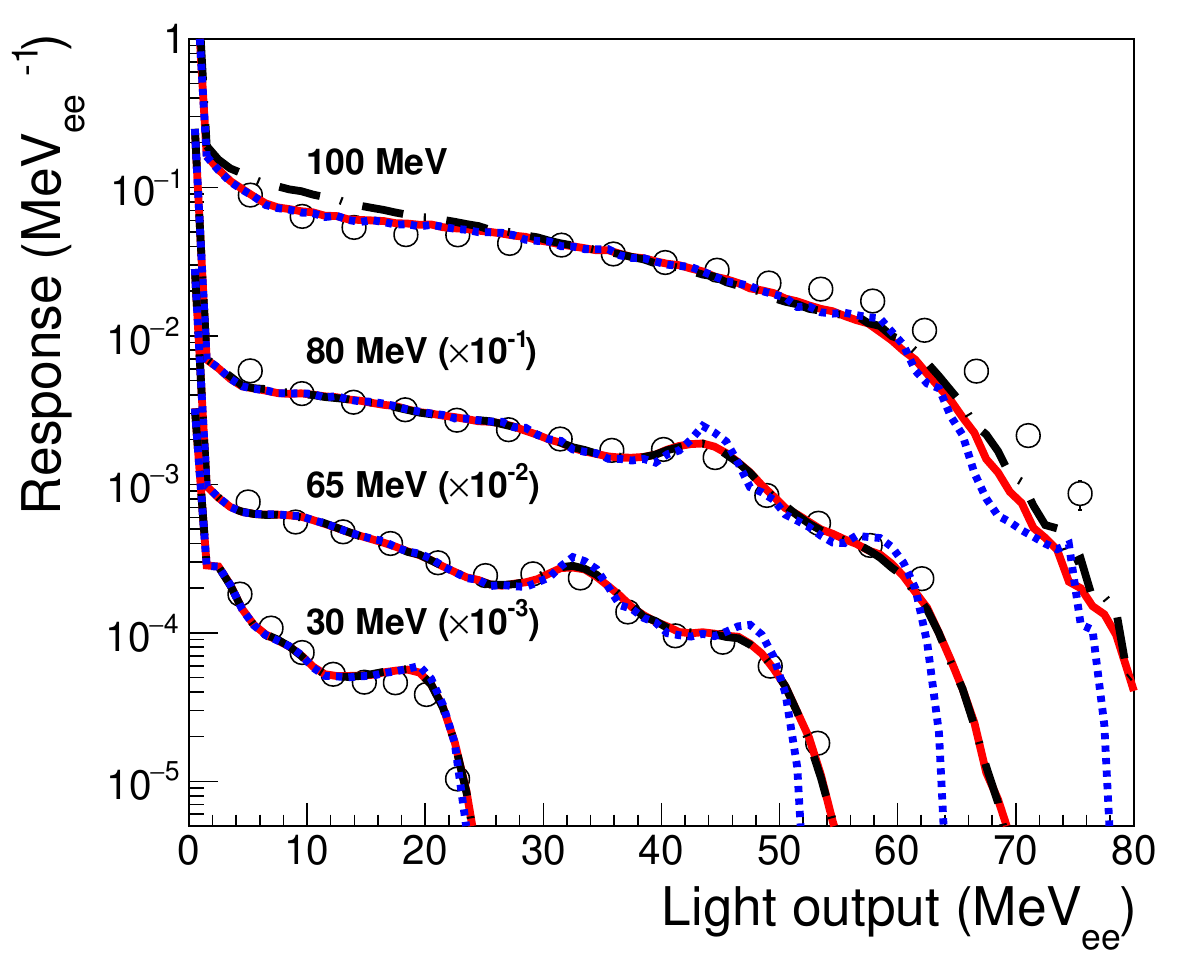}
\end{center}
\caption{Response functions of neutrons from experimental data (open circles) and neuSIM4 simulations (lines) at different energies. The experimental data are for $^{48}$Ca + $^{64}$Ni reactions at 140 MeV/u~\cite{teh-1} using the LANA Wall. The red solid and blue dotted lines represent the response functions from neuSIM4(INCL) with and without the finite resolution effect given by Eq.~\ref{eq:error}, respectively. The black dot-dashed lines represent the response function from neuSIM4(NxQMD) with the resolution effect included.}
\label{fig:lana_resp}
\end{figure}

The procedure described in Sec.~\ref{sec:scinful-geant4} is repeated for the geometry of LANA. In this work, only the front LANA wall is used. The point neutron source is located at the nominal target position, and neutrons at different energies uniformly distributed in a cone are sent to LANA. When the neutrons hit the scintillation material, light is produced by secondary charged particles. The lights are reflected on the boundary between scintillator and Pyrex container, and, finally reach the PMTs attached at both ends of each cuboid. The creation and the propagation of the lights can be simulated by utilizing optical photon processes in GEANT4 toolkit. However, because of large number of optical photons produced in NE213 scintillator, 12,000 per MeV$_{\rm{ee}}$ and the length of the scintillor bar, the calculation time increases enormously and is not sustainable. Instead of implementing optical photon processes, we incorporate a resolution function of the scintillation lights~\cite{dietze-2} taking into account the detector geometry, the statistical uncertainty for the creation of photoelectrons on the PMT photocathode and electrical noise. As a result, the resolution of the scintillation light is considered as a Gaussian distribution with full width at half maximum (FWHM) given by

\begin{equation}
{\frac{dL_{tot}}{L_{tot}}} = \sqrt{A^{2} + {\frac{B^{2}}{L_{tot}}} + {\frac{C^{2}}{L_{tot}^{2}}}},
\label{eq:error}
\end{equation}
where the first term with coefficient $A$ represents the resolution effect for a given detector geometry, the second term with coefficient $B$ is from statistical uncertainty due to fluctuation in the number of created photoelectrons, and the last term with coefficient $C$ reflects the electrical noise~\cite{dietze-2}. For this work $A$, $B$, and $C$ are treated as free fitting parameters. We run simulations with various combinations of $A$, $B$, and $C$ and compared to data shown in Fig.~\ref{fig:lana_resp}. The best fit values from the least $\chi^2$ fitting for these parameters are $A$ = 0.15$\pm$0.03, $B$ = 0.15$\pm$0.04, and $C$ = 0.02$\pm$0.01. 

Figure~\ref{fig:lana_resp} shows the response functions of neutrons from the experimental data compared with simulations by neuSIM4(INCL) (red solid lines) and neuSIM4(NxQMD) (black dash-dotted lines) at several energies from 30 to 200 MeV. The experimental data in Fig.~\ref{fig:lana_resp} are obtained from the reaction of $^{48}$Ca + $^{64}$Ni at 140 MeV/u by the HiRA collaboration~\cite{fanurs-thesis}. This experiment is designed to detect neutrons from heavy-ion collisions and does not have the ability to count the number of emitted neutrons from the reactions to the detector. Thus, the experimental data alone do not provide absolute efficiency measurements. Nonetheless, one can compare the shapes of the response functions using a normalization constant obtained by integrating a flat region of the response function typically from 25 MeV$_{\rm{ee}}$ to 60 MeV$_{\rm{ee}}$ for high energy neutrons with $E_{n}$ $\geq$ 100 MeV, and from 3 MeV$_{\rm{ee}}$ to below the $^{12}$C(n,d) peak for low energy neutrons. For neutron energies up to 80 MeV, only the low energy physics model KSCIN is used in the simulations. Thus, the neuSIM4(INCL) and neuSIM4(NxQMD) results shown in Fig.~\ref{fig:lana_resp} are exactly the same. The calculations agree with the experimental data reasonably well, especially when resolution effects (as described in Eq.~\ref{eq:error}) are included. Without the resolution effects (blue dotted lines), the peaks are slightly sharper and the fall-off at high light output is steeper. The $\chi^{2}$/ndf decreases from 7.8 to 4.5 when the effect of the resolution is included in the simulations. For neutron energy above 80 MeV the agreement with data is poor, especially at high light output where there is no data and statistics from simulations are sparse due to sharp drop in the response. The effect of including noise resolution is minimal at high energy. Nonetheless, the noise resolution correction is included regardless of the incident neutron energy. For neutron energies above 120 MeV, the neuSIM4(INCL) simulations provide surprisingly better agreement with the experimental data than the results from neuSIM4(NxQMD) as seen in Figs.~\ref{fig:cylinder_resp} and ~\ref{fig:lana_resp}. We do not understand the reason. It could be that the data in Fig.~\ref{fig:cylinder_resp} and \ref{fig:cylinder_eff}  were obtained with small cylindrical neutron detectors. In Fig.~\ref{fig:lana_resp}, we compare calculations to data from a large neutron wall array. 


\begin{figure}[t!]
\begin{center}
\includegraphics[width=0.95\linewidth]{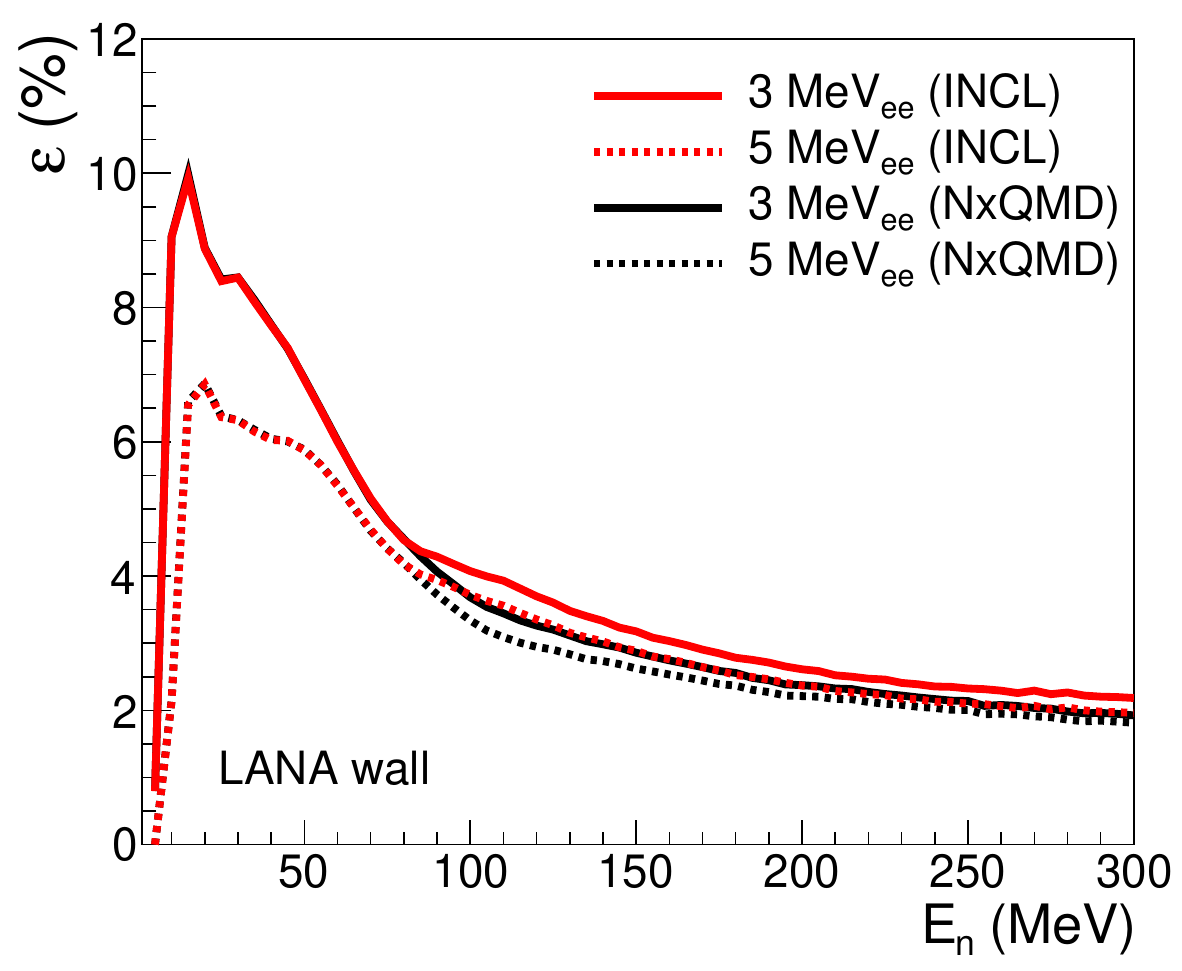}
\end{center}
\caption{Neutron detection efficiencies of LANA as a function of energy, determined by simulations using neuSIM4(INCL) (red solid and red dotted lines) and neuSIM4(NxQMD) (black solid and black dotted lines). The solid and dotted lines are obtained with light output thresholds, $L_{tot}^{thr}$, of 3 and 5 MeV$_{\rm{ee}}$, respectively.}
\label{fig:lana_eff}
\end{figure}

\subsection{Simulated neutron detection efficiency for LANA} \label{sec:lana-sim-eff}

Neutron detection efficiency is obtained by integrating the response functions, presented in Fig.~\ref{fig:lana_resp}, from the light output threshold value of $L_{tot}^{thr}$ to infinity for a given neutron energy and normalizing it by the number of incident neutrons. Figure~\ref{fig:lana_eff} shows the neutron detection efficiency of the LANA wall as a function of neutron energy.
Experimentally, neutrons from a source or a special accelerator are used to determine the neutron detection efficiency of a detector. In such a setup, the number of neutrons incident on the detector can be determined. However, in the experiment that provided the light output response in Figure~\ref{fig:lana_resp}, the detected neutrons came from heavy ion collisions, and we could not determine the exact number of incident neutrons. Thus, we are not able to determine the absolute neutron efficiencies and their associated uncertainties. We note that the accuracy of the neutron efficiency determination does not depend on the complete agreement in the simulation of the light response function to data, especially at high light output region where the response drops exponentially. We adopt uncertainties on the simulation efficiencies of 15\% above 80 MeV and 10\% for neutron energy below 80 MeV from Refs.~\cite{Iwamoto-1, satoh-1, satoh-2}. As will be discussed in the next section, uncertainties in the efficiencies also depend on the number of neutrons hitting the detectors simultaneously. More study on the uncertainties will be needed when the efficiency is applied to experimental data.

In Fig.~\ref{fig:lana_eff}, the red curves depict the results obtained from neuSIM4(INCL), while the black curves represent results from neuSIM4(NxQMD). The efficiencies from the two options start to deviate around 80 MeV when calculations from the high energy physics model start to contribute. Efficiencies obtained from NxQMD are smaller than those from INCL but are within the uncertainties of the simulation efficiencies. The solid curves and dotted curves correspond to $L_{tot}^{thr}$ values of 3 and 5 MeV$_{\rm{ee}}$, respectively. Detection efficiencies are larger with lower light output threshold values. The largest difference between results from the two different light output thresholds occurs when the neutron energy is around 20 MeV. After that, the differences start to decrease with neutron energy. It is reduced to less than $\sim$0.5\% around 100 MeV and becomes even smaller at higher energies.

\begin{figure*}[t!]
\begin{center}
\includegraphics[width=0.95\linewidth]{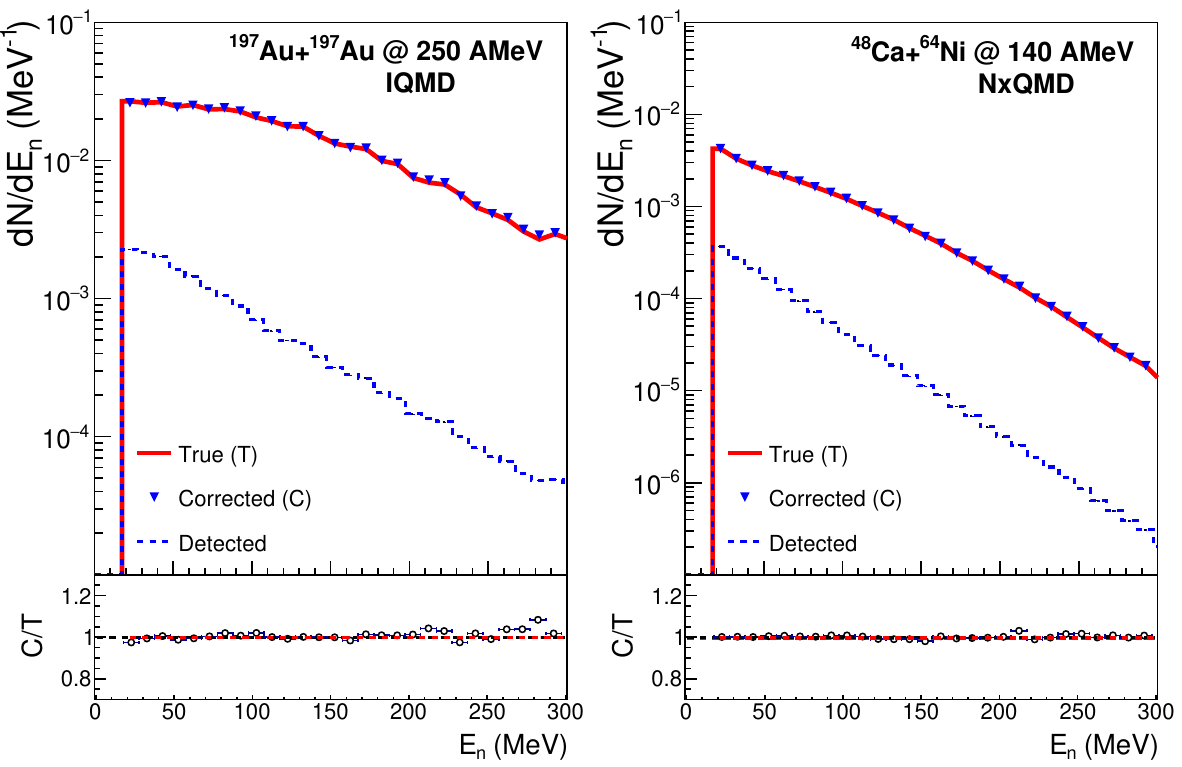}
\end{center}
\caption{Comparison of the true neutron spectra (T) with the efficiency corrected neutron spectra (C) from the physics event generators using the LANA Wall setup. The left and right panels show the simulation results for $^{197}$Au + $^{197}$Au at 250 MeV/u by IQMD and $^{48}$Ca + $^{64}$Ni at 140 MeV/u by NxQMD, respectively. The light output threshold for the simulation is set to 3 MeV$_{\rm{ee}}$.}
\label{fig:lana_spectra}
\end{figure*}

To validate the determined detection efficiency for LANA, we conducted a closure test using two transport model event generators. A closure test is a procedure used in statistical analysis to verify the consistency and accuracy of our simulation by comparing the corrected spectra with the true spectra from the event generator. For a closure test, it is not important that an accurate simulation code is used. For this test, the data are generated with different QMD codes mainly due to availability of  events from existing simulations or existing codes available to the authors.  In this study, we used the events from two collision systems: $^{197}$Au + $^{197}$Au at 250 MeV/u generated by the Isospin-dependent Quantum Molecular Dynamics (IQMD) model, and $^{48}$Ca + $^{64}$Ni at 140 MeV/u using NxQMD which is at our disposal. The former collision system is selected as a representative of events with high-neutron-multiplicity and the latter system is chosen because experimental data exist. In the simulations, LANA has the same configuration as the experimental setup. 

Neutrons incident on LANA interact with the detector according to the same neutron scattering physics we describe earlier including folding in the electronic resolution of the light outputs in accordance with Eq.~\ref{eq:error}. The resulting neutron spectra is labeled as "detected" (blue dotted lines) in Figure~\ref{fig:lana_spectra}. Then the neutron efficiency described in this work is applied to obtain the "corrected" spectra (blue inverted triangles) which are compared to the true neutron spectra (red solid lines) obtained directly from the physics event generators for both systems. The corrected spectra practically overlap with the true spectra indicating our correction procedure has been applied correctly. For the simulation $L_{tot}^{thr}$ is set to 3 MeV$_{\rm{ee}}$. Increasing $L_{tot}^{thr}$ to 5 MeV$_{\rm{ee}}$ gives the same results as long as the same condition applied to the data is also applied to the simulation. The bottom panels of Fig.~\ref{fig:lana_spectra} show the ratios of the corrected to the true distributions, $C/T$, of neutrons from the simulation. The black dashed horizontal line is the unity line at C/T=1. Deviations of C/T from unity for the low energy reactions in the right panel are much less than the high energy reactions. The larger uncertainties may be due to higher multiplicity neutrons emitted from the the higher energy reactions with heavier nuclei. The uncertainties are less than 3\% for neutrons below 200 MeV. This is very good for the experiment $^{48}$Ca + $^{64}$Ni at 140 MeV/u where emitted neutron energies have been measured up to around 200 MeV. For higher energies, one needs to employ more sophisticated methods to correct for multiple neutrons hitting LANA. 


\section{Summary} \label{sec:sum} 


We developed and implemented two new physics models in GEANT-4 to simulate low- and high-energy neutron interactions with organic scintillators. The core-code, `neuSIM4', features flexible detector geometry including complicated setups with multiple neutron detectors. It can simulate the performance of neutrons with energy up to 110 MeV using the KSCIN physics model. Previously, only one low energy physics model in GEANT4 was accurate up to 20 MeV, partly due to the lack of extensive databases of n+C reaction channels in other codes. In contrast, databases of neutron induced reaction cross-sections on H and C nuclei are incorporated in neuSIM4 to ensure accuracy of the simulations. For neutron energies greater than 110 MeV, a quantum molecular dynamic model, NxQMD, is used to successfully describe high energy neutron interactions. Notably, we find that an existing cascade physics model (INCL) in GEANT4 performs similarly to NxQMD but with significantly reduced computing time. Our program provides two options: neuSIM4(NxQMD) and neuSIM4(INCL). 
After validation with data, both new codes are applied to calculate the performance of a 2 $\times$ 2 m$^{2}$ neutron detector array with 25 scintillation bars (LANA). Differences between efficiencies from the two options are small. 
The maximum value of the neutron efficiency for LANA is predicted to be around 10\% at neutron energy of 20 MeV and decreases to about 3\% for 300 MeV neutrons. Using the calculated efficiency for LANA, the experimental neutron energy spectrum from $^{48}$Ca + $^{64}$Ni collisions at 140 MeV/u can be reconstructed. This will be used to investigate the symmetry energy and in-medium properties of isospin asymmetric nuclear matter. The current codes, neuSIM4(NxQMD) and neuSIM4(INCL) will play important roles in analyzing neutron data measured by modern detectors with a variety of geometries~\cite{shim-1}.

\section*{Code availability}
Eventually the code will be available for download at Github. In the meantime, it is available upon request to the corresponding author.
 
\section{Acknowledgments}
The authors are grateful to Dr. Daiki Satoh at the JAEA for providing answers and insights to our many questions about SCINFUL-NxQMD (known as SCINFUL-QMD in the literature) and SCINFUL-PHITS codes. 
This work was supported by the National Research Foundation funded by MSIT of Korea (Grant No. 2018R1A5A1025563) and U.S. National Science Foundation Grant No. PHY-2209145 and PHY-2110218.



\begin{thebibliography}{99}
\bibitem{erler-1} J. Erler, $et~al.$, Nature {\bf 486} (2012) 509.
\bibitem{lynch-1} W.G. Lynch, M.B. Tsang, Phys. Lett. B {\bf 830} (2022) 137098. 
\bibitem{russotto-1} P. Russotto, $et~al.$, Riv. Nuovo Cimento {\bf 46} (2023) 1.
\url{https://doi.org/10.1007/s40766-023-00039-4}. 
\bibitem{schatz-1} H. Schatz, $et~al.$, J. Phys. G {\bf 49} (2022) 110502.
\bibitem{wallace-1} M.S. Wallace, $et~al.$, Nucl. Instr. and Meth. A {\bf 583} (2007) 302-312 
\bibitem{pollacco-1} E. Pollacco $et~al.$, Eur. Phys. J. A {\bf 25}, (2005) 287. 
\bibitem{shim-1} H. Shim $et~al.$, Nucl. Instr. and Meth. A {\bf 927} (2019) 280. 
\bibitem{brun-1} R. Brun $et~al.$, CERN-DD-EE-84-1 (1987), ``GEANT3" 
\bibitem{fluka-1} B{\"o}hlen T.T. $et~al.$, Nucl. Data Sheets, {\bf 120} (2014) 211, ``The FLUKA code: Developments and challenges for high energy and medical applications" 
\bibitem{dietze-1} G. Dietze, "NRESP4 and NEFF4, Monte-Carlo codes for the calculation of neutron response functions and detection efficiencies for NE-213 scintillation detectors", Neutronendosimetrie: PTB-Bericht, Bundesanst, 1982, 
\url{https://books.google.co.kr/books?id=Tw1PmgEACAAJ}. 
\bibitem{dickens-1} J.K. Dickens, "SCINFUL: A Monte-Carlo based computer program to determine a scintillator full energy response to neutron detection for $E_{n}$ between 0.1 and 80 MeV: User's manual and FORTRAN program listing", 1988, 
\url{https://www.osti.gov/biblio/5178779}. 
\bibitem{agostinelli-1} S. Agostinelli $et~al.$, Nucl. Instr. and Meth. A {\bf 506} (2003) 250. 
\bibitem{kohley-1} Z. Kohley $et~al.$, Nucl. Instr. and Meth. A {\bf 682} (2012) 59. 
\bibitem{hermann-1} H. Wolter $et~al.$, Prog. Part. Nucl. Phys. {\bf 125} (2022) 103962.
\bibitem{coupland-1} Daniel David Schechtman Coupland, (2013) "Probing the nuclear symmetry energy with heavy ion collisions", Doctoral dissertation, Michigan State University.
\bibitem{satoh-1} D. Satoh $et~al.$, J. Nucl. Sci. Tech. 2 (2002) 657. 
\bibitem{satoh-com} D. Satoh $et~al.$, JAEA-Data/Code, 2006-023, 1-43, (2006). 
\bibitem{niita-1} K. Niita $et~al.$, Phys. Rev. C {\bf 52} (1995) 2620. 
\bibitem{mancusi-1} D. Mancusi $et~al.$, Phys. Rev. C {\bf 90}, (2014) 054602. 
\bibitem{satoh-4} D. Satoh and T. Sato, J. Nucl. Sci. Tech. {\bf 59} (2022) 1047. 
\bibitem{fanurs-thesis} Fanurs Chi-En Teh, (2023) "Constraining nucleon effective mass splitting using neutron and proton observables from heavy-ion collisions", Doctoral dissertation, Michigan State University.
\bibitem{antolkovic-1} B. Antolkovic $et~al.$, Nucl. Phys. A {\bf 394} (1983) 87.
\bibitem{subramanian-1} T. S. Subramanian $et~al.$, Phys. Rev. C {\bf 28} (1983) 521.
\bibitem{mcmillan-1} Edwin M. Mcmillan $et~al.$, Phys. Rev. {\bf 73} (1948) 262.
\bibitem{kellogg-1} D. A. Kellogg, Phys. Rev. {\bf 90} (1953) 224.
\bibitem{chiba-1} S. Chiba $et~al.$, J. Nucl. Sci. Tech. {\bf 33} (1996) 654. 
\bibitem{monning-1} F. M{\"o}nnig, H. Schopper, 1.1.2 Total cross sections of neutrons on protons and nuclei: datasheet from Landolt-B{\"o}rnstein - Group I elementary particles, nuclei and atoms Volume 7: ``Elastic and charge exchange scattering of elementary particles" in SpringerMaterials. Heidelberg (Berlin): Springer-Verlag; 1973. 
\bibitem{satoh-3} D. Satoh $et~al.$, Radiat. Prot. Dosim. {\bf 126} (2007) 555. 
\bibitem{nakao-1} N. Nakao $et~al.$, Nucl. Instr. and Meth. A {\bf 463} (2001) 275. 
\bibitem{mendoza-1} E. Mendoza $et~al.$, IEEE Tran. Nucl. Sci. {\bf 61} (2014) 2357. 
\bibitem{thulliez-1} L. Thulliez $et~al.$, Nucl. Instr. and Meth. A {\bf 1027} (2022) 166187. 
\bibitem{garcia-1} A.R. Garcia $et~al.$, Nucl. Instr. and Meth. A {\bf 868} (2017) 73. 
\bibitem{boudard-1} A. Boudard $et~al.$, Phys. Rev. C {\bf 87} (2013) 014606. 
\bibitem{boudard-2} A. Boudard $et~al.$, Phys. Rev. C {\bf 66} (2002) 044615. 
\bibitem{folger-1} G. Folger $et~al.$, Eur. Phys. J. A {\bf 21}, (2004) 407. 
\bibitem{gurthrie-1} M. P. Guthrie $et~al.$, Nucl. Instr. Meth. {\bf 66}, (1968) 29. 
\bibitem{koi-1} T. Koi, New native QMD code in Geant4, in: Proc. SNA + MC2010: Joint international conference on supercomputing in nuclear applications + Monte Carlo 2010 Tokyo, 2010, 1630. 
\bibitem{satoh-2} D. Satoh $et~al.$, J. Nucl. Sci. Tech. {\bf 43} (2006) 714. 
\bibitem{zecher-1} P.D. Zecher $et~al.$, Nucl. Instr. and Meth. A {\bf 401} (1997) 329. 
\bibitem{zhu-1} K. Zhu $et~al.$, Nucl. Instr. and Meth. A {\bf 967} (2020) 163826. 
\bibitem{teh-1} F.C.E. Teh $et~al.$, IEEE Tran. Nucl. Sci. {\bf 68} (2021) 2294. 
\bibitem{dietze-2} G. Dietze and H. Klein, Nucl. Instr. and Meth. {\bf 193} (1982) 549. 
\bibitem{Iwamoto-1} Y. Iwamoto $et~al.$, Nucl. Instr. and Meth. A {\bf 804} (2015) 50. 

\end{thebibliography}
\end{document}